\documentclass[aps,prl,twocolumn,superscriptaddress]{revtex4-1}
\usepackage{graphicx}
\usepackage{epsfig}
\usepackage{amsmath,amsfonts,amsthm}
\usepackage{amssymb}
\usepackage{times}
\usepackage{braket}
\usepackage{units}
\usepackage{multirow}
\usepackage{mathrsfs}
\usepackage{theorem}
\usepackage{bm}
\usepackage{threeparttable}
\usepackage{url}
\usepackage[T1]{fontenc}
\usepackage{csquotes}
\usepackage{natbib}
\usepackage{float}
\usepackage{appendix}
\usepackage[colorlinks=true,linkcolor=blue,citecolor=blue,urlcolor=blue]{hyperref}
\usepackage{bookmark}
\usepackage{soul}
\MakeOuterQuote{"}
\usepackage{dcolumn}

\newcommand{\Tr}{\operatorname{Tr}}
\def\>{\rangle}
\def\<{\langle}
\def\bb{\langle\!\langle}
\def\kk{\rangle\!\rangle}
\newcommand{\map}[1]{\mathcal{#1}}

\begin{document}

\title{Experimental transmission of quantum information using a superposition of causal orders}
\author{Yu Guo}
\affiliation{CAS Key Laboratory of Quantum Information, University of Science and Technology of China, Hefei, 230026, People's Republic of China}
\affiliation{CAS Center For Excellence in Quantum Information and Quantum Physics, University of Science and Technology of China, Hefei 230026, P.R. China}
\author{Xiao-Min Hu}
\affiliation{CAS Key Laboratory of Quantum Information, University of Science and Technology of China, Hefei, 230026, People's Republic of China}
\affiliation{CAS Center For Excellence in Quantum Information and Quantum Physics, University of Science and Technology of China, Hefei 230026, P.R. China}
\author{Zhi-Bo Hou}
\affiliation{CAS Key Laboratory of Quantum Information, University of Science and Technology of China, Hefei, 230026, People's Republic of China}
\affiliation{CAS Center For Excellence in Quantum Information and Quantum Physics, University of Science and Technology of China, Hefei 230026, P.R. China}
\author{Huan Cao}
\affiliation{CAS Key Laboratory of Quantum Information, University of Science and Technology of China, Hefei, 230026, People's Republic of China}
\affiliation{CAS Center For Excellence in Quantum Information and Quantum Physics, University of Science and Technology of China, Hefei 230026, P.R. China}
\author{Jin-Ming Cui}
\affiliation{CAS Key Laboratory of Quantum Information, University of Science and Technology of China, Hefei, 230026, People's Republic of China}
\affiliation{CAS Center For Excellence in Quantum Information and Quantum Physics, University of Science and Technology of China, Hefei 230026, P.R. China}
\author{Bi-Heng Liu}
\email{bhliu@ustc.edu.cn}
\affiliation{CAS Key Laboratory of Quantum Information, University of Science and Technology of China, Hefei, 230026, People's Republic of China}
\affiliation{CAS Center For Excellence in Quantum Information and Quantum Physics, University of Science and Technology of China, Hefei 230026, P.R. China}
\author{Yun-Feng Huang}
\affiliation{CAS Key Laboratory of Quantum Information, University of Science and Technology of China, Hefei, 230026, People's Republic of China}
\affiliation{CAS Center For Excellence in Quantum Information and Quantum Physics, University of Science and Technology of China, Hefei 230026, P.R. China}
\author{Chuan-Feng Li}
\email{cfli@ustc.edu.cn}
\affiliation{CAS Key Laboratory of Quantum Information, University of Science and Technology of China, Hefei, 230026, People's Republic of China}
\affiliation{CAS Center For Excellence in Quantum Information and Quantum Physics, University of Science and Technology of China, Hefei 230026, P.R. China}
\author{Guang-Can Guo}
\affiliation{CAS Key Laboratory of Quantum Information, University of Science and Technology of China, Hefei, 230026, People's Republic of China}
\affiliation{CAS Center For Excellence in Quantum Information and Quantum Physics, University of Science and Technology of China, Hefei 230026, P.R. China}
\author{Giulio Chiribella}
\email{giulio.chiribella@cs.ox.ac.uk}
\affiliation{Department of Computer Science, The University of Hong Kong, Pokfulam Road, Hong Kong, P.R. China}
\affiliation{Department of Computer Science, University of Oxford, Parks Road, Oxford, UK}
\begin{abstract}
Communication  in a network generally takes place through a sequence of intermediate nodes connected by communication channels.   In the standard theory of communication, it is  assumed that the communication network is embedded in a classical spacetime, where the relative order of different nodes is well-defined. In principle, a quantum theory of spacetime could allow the  order of the intermediate points between sender and receiver to be in a coherent superposition. Here we experimentally realise a table-top simulation of this exotic possibility on a photonic system, demonstrating high-fidelity transmission of quantum information over  two noisy  channels arranged in a superposition of two alternative causal  orders.
\end{abstract}

\date{\today}

\maketitle
\paragraph {Introduction.---} Communication from a sender to a receiver generally takes place through a series of  intermediate nodes. For example, an email sent over the internet is generally  relayed by a sequence of servers before  landing in the receiver's inbox.
Classically, the order of the intermediate nodes is always well-defined.  The communication network  is embedded in a classical spacetime where the causal relations between points are fixed.   In quantum theory, instead, the superposition principle  suggests that there may exists scenarios where spacetime itself  is in a  superposition of alternative configurations \cite{rovelli2004quantum,hardy2007towards}. A communication network embedded in a quantum spacetime could  give rise to new scenarios where the communication channels act  in a quantum superposition of orders  \cite{chiribella2013quantumcomputations,chiribella2012perfect} or in some other form of indefinite order \cite{oreshkov2012quantum}.

The extension of communication theory to scenarios where quantum channels act  in a  superposition of orders was recently addressed in a series of theoretical works  \cite{ebler2018enhanced,salek2018quantum,chiribella2018indefinite,chiribella2019quantum,procopio2019communication}. These works demonstrated  various advantages over the standard communication model of quantum  Shannon theory \cite{wilde2013quantum}, where the order of communication channels is well-defined. For example, Ref. \cite{ebler2018enhanced} showed that two completely depolarising channels acting in a superposition of two orders can transmit a non-zero amount of classical information, whereas in the standard model  they would completely block any kind of information.  Similar advantages arise in the transmission of quantum information  \cite{salek2018quantum}, sometimes leading to a complete removal of the noise \cite{chiribella2018indefinite}. To what extent these advantages  are specific to superpositions of  causal orders, rather than being generic to other forms of coherent superpositions of communication protocols,   is currently a matter of debate \cite{salek2018quantum,abbott2018communication,chiribella2019quantum,guerin2018communication,loizeau2019channel,kristjansson2019resource}. Nevertheless, the advantages  of the superposition of orders suggest that having access to quantum superpositions of spacetimes could have major  consequences on the power of   quantum communication networks.

The search for experimental evidence of quantum spacetimes has attracted increasing interest in recent years \cite{hossenfelder2017experimental,bose2017spin,marletto2017gravitationally,christodoulou2019possibility}. Still, no direct observation has   been possible so far. An alternative approach  is to simulate the superposition of spacetimes through the already accessible physics taking place in a fixed, classical spacetime.  The action  of two communication channels connected in a quantum superposition of causal orders can be reproduced by setting up a mechanism that  routes photons  through two optical  devices and controls the order in which the devices are visited \cite{procopio2015experimental,rubino2017experimental,goswami2018indefinite}.  In this way, it is possible to simulate   quantum communication networks where the communication channels are embedded in a superposition of spacetimes, and to witness the advantages of the corresponding communication model.

Here we experimentally demonstrate the possibility of  high-fidelity transmission of both classical and quantum information through noisy channels in a superposition of causal orders \cite{ebler2018enhanced,salek2018quantum,chiribella2018indefinite}. Our results offer a glance at exotic communication scenarios that could  arise in quantum spacetimes, and, at the same time, contribute to the development of  a new technology of coherent control over multiple transmission lines, with  applications beyond the study of communication protocols with superpositions of orders.  Indeed,  the ability to coherently route photons through multiple optical  devices  enables not only communication in  a superposition of  causal orders, but also a variety of new communication protocols involving superpositions of alternative communication channels \cite{gisin2005error,abbott2018communication,chiribella2019quantum},  superpositions of alternative directions of communication \cite{delsanto2018two}, and  superpositions of alternative encoding/decoding operations \cite{guerin2018communication}. The high level of accuracy  achieved by our setup   can also  benefit the realization of these protocols, thereby contributing to the exploration of  a broader class of quantum communication networks where the communication takes place in superpositions of alternative configurations.

\paragraph {Background.---}In a standard  communication scenario, a sender transmits  messages to a receiver by encoding them in the internal state of a quantum particle and then sending it to the receiver through a sequence of noisy channels. The action of a generic quantum channel $\mathcal E$  on a given quantum input state $\rho$ is conveniently expressed in the Kraus representation $\mathcal{E}(\rho)=\sum_i E_i \rho E_i^{\dagger}$,
where $\{E_i\}$ are linear operators satisfying $\sum_iE_iE_i^{\dagger}=I$, $I$  being  the identity operator.

The ability of a channel to transmit classical and quantum information is quantified by its classical capacity $\emph{C}$ and quantum capacity $\emph{Q}$, respectively. The classical capacity $\emph{C}$ is  the maximum number of bits that can be reliably transmitted per channel use, in the limit of asymptotically many channel uses \cite{holevo1998capacity,schumacher1997sending}.
A lower bound one the classical capacity is provided by the one-shot accessible information
\begin{equation}\label{equ2}
C_1=\max_{\{p_x,\rho_x\}}  \max_{\{  P_y\}}H(X:Y)
\end{equation}
where the maximum is  over all possible input ensembles$\{p_x,\rho_x\}$ and output measurements $\{  P_y\}$,  and $H(X:Y)$ is the mutual information between $x$ and $y$.

The quantum capacity $\emph{Q}$ is the maximum number of qubits that can be reliably transmitted per channel use, again in the limit of asymptotically many uses  \cite{Lloyd1997,Shor2002,Devetak2005}.  A lower bound to the quantum capacity is the one-shot coherent information
\begin{equation}\label{equ3}
{Q}_1=\max\limits_{\rho} \, \mathcal{I}_c(\rho)
\end{equation}
where $\mathcal{I}_c  (\rho): = S[\mathcal{E}(\rho)]-S_e(\rho,\mathcal{E})$ is the coherent information of the state $\rho$~\cite{schumacher1996quantum}, defined in terms of the von Neumann entropy $S(\rho):=-\Tr[\rho \log_2\rho]$, and of the entropy exchange  $S_{e}(\rho,\mathcal{E}):=S[(\map{I}_R\otimes\mathcal{E})(|\Psi_{
\rho}\rangle\langle\Psi_{\rho}|)]$, $|\Psi_{\rho}\rangle$ being any purification of $\rho$ using a reference quantum system $R$.

\paragraph {Communication networks and superposition of orders.---}In a network scenario,  the communication between a sender and a receiver proceeds through a sequence of intermediate nodes, located at different spacetime points and connected by  communication channels.    Here we consider the case of $n=2$ channels, $\map E$ and $\map F$,   the former  connecting spacetime points $P$ and  $P'$, and the latter connecting spacetime points $Q$ and $Q'$.

In the standard setting,  the causal relations among spacetime points are  well-defined, and so is the order of the intermediate nodes between the sender and receiver, located at spacetime points $S$ and $R$, respectively.  For example, one can have the order $S\preceq P\preceq P'\preceq Q \preceq Q'\preceq R$, indicating that signals can be transmitted from $S$ to $P$, from $P$ to $P'$, and then to $Q$ and to $Q'$.    Assuming  that no noise takes place in the transmission from  $S$ to $P$, $P'$ to $Q$,  and $Q$ to $R$, this configuration leads to the overall channel  $\map F \map E$ connecting the sender and receiver.     In another spacetime configuration, one could    have the order   $S\preceq Q\preceq Q' \preceq P\preceq P'\preceq R$, corresponding to the channel  $\map E\map F$.   In either configurations, the sender and receiver are assumed to know the structure of spacetime, and therefore to know whether the overall channel between them is $\map E \map F$ or $\map F \map E$.

New possibilities arise when the background spacetime is treated quantum mechanically \cite{salek2018quantum}. One can associate the basic configurations $S\preceq P\preceq P'\preceq Q \preceq Q'\preceq R$  and $S\preceq Q\preceq Q' \preceq P\preceq P'\preceq R$ to two orthogonal states    $|0\>$ and  $|1\> $,  forming a basis for an effective two-dimensional quantum system, called the {\em order qubit}.  The order qubit can be interpreted as a coarse-grained description of a quantum spacetime in which the communication network is embedded. In this scenario, the basis states $|0\>$ and $|1\>$ represent two distinct semi-classical states of the gravitational field. Coherent superpositions semi-classical states arise generically across  various theories of quantum gravity \cite{bose2017spin,marletto2017gravitationally,christodoulou2019possibility}.


The insertion of  two  quantum channels $\map E$ and $\map F$ into a quantum spacetime  in the state $\omega$ can be  described by the  quantum SWITCH (QS) transformation \cite{chiribella2013quantumcomputations,chiribella2012perfect} $\map S_\omega  :  (\map E, \map F) \mapsto \map S_\omega (\map E, \map F)$, defined as
\begin{align}\label{switchdef1}
\map S_{\omega} (\map E, \map F)\left(\rho\right):= \sum_{i,j} W_{ij} \left(\rho \otimes \omega \right) W_{ij}^\dagger,  \end{align}
with
\begin{align}\label{switchdef2}
W_{ij}:=E_i F_j \otimes |0\>\<0| + F_j  E_i \otimes   |1\>\<1|  \,,
\end{align}
where $\{E_i\}$ and $\{F_j\}$ are the Kraus operators of $\map E$ and $\map F$.
The quantum channel  (\ref{switchdef1}) can be reproduced in the ordinary, classically-well defined spacetime, using {\em e.g.} photonic systems \cite{procopio2015experimental,rubino2017experimental,goswami2018indefinite}.

It is worth noting that quantum channel (\ref{switchdef1})  is independent of the choice of Kraus operators $\{E_i\}$ and $\{F_j\}$.  Physically, this means  that, in principle, its realization does not require access to the environment of the communication devices. This is not the case, however, in all the existing  implementations of the channel $S_\omega  (\map E,\map F)$ (including the present one),  where access to the environments of both channels $\map E$ and $\map F$ is essential for producing the superposition of causal orders while operating in a classical spacetime \cite{oreshkov2018whereabouts,chiribella2019quantum}.

The interference between the two alternative orders provides  advantages  over standard quantum Shannon theory, enabling communication through channels that individually block information \cite{ebler2018enhanced,salek2018quantum,chiribella2018indefinite,chiribella2019quantum}.  While these advantages crucially rely on the order qubit being used to assist the decoding,  it is worth stressing that  they do not use the order qubit as a way to bypass the given communication channels. Indeed, the transformation of resources $\map S_\omega  :  (\map E,\map F) \mapsto  \map S_\omega (\map E,\map F)$ does not give the sender and the receiver a way to transmit information independently of the channels  $\map E$ and $\map F$ \cite{salek2018quantum,kristjansson2019resource}.

The advantages in Refs. \cite{ebler2018enhanced,salek2018quantum,chiribella2018indefinite,chiribella2019quantum} involve pairs of Pauli channels,  of the form $\mathcal{E}_{\vec{p}}   =  \sum_{i=0}^3 p_i\sigma_i\rho\sigma_i$   and $\mathcal{F}_{\vec{q}}  =  \sum_{i=0}^3 q_i\sigma_i\rho\sigma_i$, where $(\sigma_0,\sigma_1,\sigma_2,\sigma_3)$ are the Pauli matrices $(I,X,Y,Z)$. Suppose that the two channels $\map E_{\vec p}$ and $\map F_{\vec q}$  are combined in a superposition of orders,
with the control qubit is in the state $\omega=  |+\>\<+|$. From Equations~(\ref{switchdef1})and (\ref{switchdef2}), the resulting channel $\map S_\omega (\map E_{\vec p}, \map F_{\vec q})$  acts as
\begin{align}\label{switched}
 \mathcal{S}(\mathcal{E}_{\vec{p}},\mathcal{F}_{\vec{q}})(\rho)= r_+\map C_+(\rho)\otimes|+\rangle\langle+|+r_-\map C_-(\rho)\otimes|-\rangle\langle-| \, ,
\end{align}
where $r_+$ and $r_-$ are the probabilities defined by  $r_- : =r_{12}  +  r_{23}  + r_{13}$,  $r_{ij}:  = p_iq_j  + p_j q_i$, and $r_+ : = 1-  r_-$, and $\map C_+$ and $\map C_-$ are the Pauli channels defined by
\begin{align}\label{C+}
 \map C_+=\frac{(\sum_{i=0}^3  r_{ii}/2)\rho+\sum_{i=1}^3 \,  r_{0i} \, \sigma_i\rho\sigma_i}{r_+}
\end{align}
and
\begin{align}\label{C-}
 \map C_-=\frac{[   r_{12} \,  \sigma_3\rho\sigma_3+  r_{23}  \,\sigma_1\rho\sigma_1+  r_{31} \,   \sigma_2\rho\sigma_2]}{  r_-} \, .
\end{align}

Hence, a receiver who measures the order qubit in the Fourier basis $\{|+\>,  |-\>\}$ can separate the two channels $\map C_+$ and $\map C_-$, and adapt the decoding operations to them.

\begin{figure}[tbph]
\begin{center}
\includegraphics [width= 1\columnwidth]{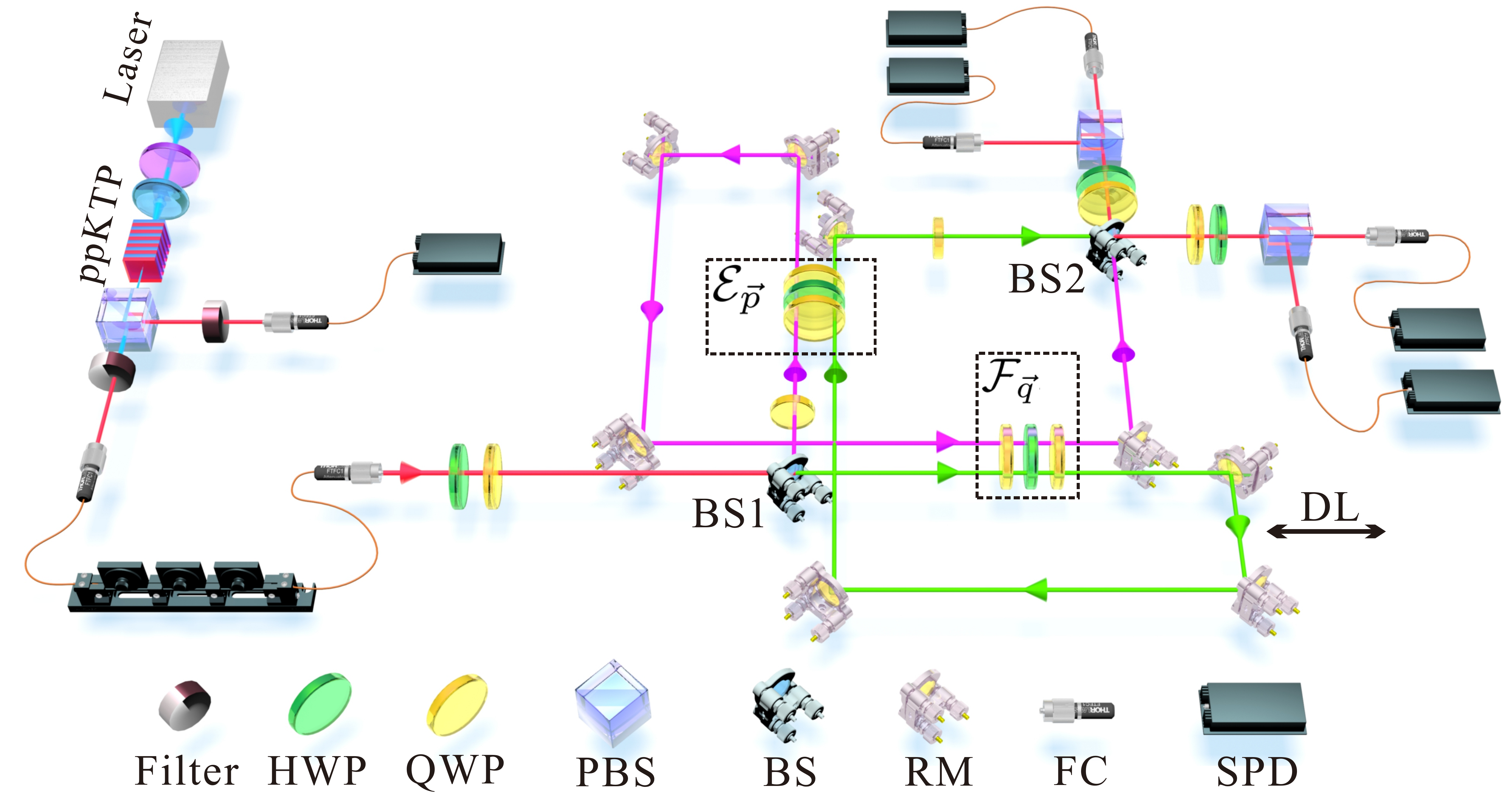}
\end{center}
\caption{Experimental setup. A cw violet laser (power is 2~mW, working at 404~mm) is incident on and pumps a type-II cut ppKTP crystal generating photon pairs of degenerate wavelength at 808nm. One of the photons acts as a trigger and the other is used to transmit information from sender to receiver. In the QS, the information is encoded in photon's polarization states while its spatial modes act as the control qubit. The Pauli channels, $\map E_{\vec p}$ and $\map F_{\vec q}$, are each composed of two QWPs and a HWP. A trombone-arm delay line and a PZT are used to set the path length and the relative phases of the interferometer. HWPs were used after BS1 and before BS2 to compensate the reflection phases introduced by the BSs. HWP: half wave plate; QWP: quarter wave plate; PBS: polarizing beam splitter; BS: beam splitter ($T/R=50/50$); RM: reflection mirror; FC: fiber coupler; SPD: single photon detector; DL: trombone-arm delay line.}
\label{fig:1}
\end{figure}


\paragraph {Experimental implementation.---}In our experiment, we demonstrated  the three communication protocols of Refs. \cite{ebler2018enhanced,salek2018quantum,chiribella2018indefinite} to a high degree of accuracy.
As shown in Fig.~\ref{fig:1}, photon pairs were generated by means of the process of spontaneous parametric down conversion. The idler photon was used as a herald and the target photon was fed into the quantum channel after encoding by the sender and then detected by the receiver. In our realisation of the switched channel  (\ref{switched}),  photonic polarization acts as the information carrying qubit, while the  spatial modes  are used as the order qubit.  Spatial modes were introduced by BS1 to switch the two channels$\mathcal{E}_{\vec{p}}$   and $\mathcal{F}_{\vec{q}}$, and BS2 was used to project the control qubit onto $|\pm\rangle_c$. We assembled two quarter wave plates (QWP) and a half wave plate (HWP) to achieve the operations $\sigma_0$, $\sigma_1$, $\sigma_2$, and $\sigma_3$. After these four operations were applied in four separate experiments, arbitrary Pauli channel $\mathcal{E}_{\vec{p}}$ can be obtained by post-processing the experimental outcomes, mixing its statistics with probabilities $\vec{p}$.

To fully characterize the action of the channel $\map S_\omega (\map E,\map F)$ on an arbitrary input state $\rho$, we performed quantum process tomography \cite{chuang1997prescription,d2001quantum,sacchi2001maximum}, where the sender prepared signal states $|H\rangle$, $|V\rangle$, $|D\rangle$, $|A\rangle$, $|R\rangle$, and $|L\rangle$ and the observables $\sigma_1$, $\sigma_2$, and $\sigma_3$ were measured by the receiver.  A generic channel $\map E$ can be reconstructed from the matrix $  \gamma_{ij}$ in  the expression   $\map E (\rho)=\sum_{ij}\gamma_{ij} \, \sigma_i \rho \sigma_j $.

\paragraph {Quantum communication with entanglement-breaking channels.---}Consider a bit flip channel $\mathcal{B}_s(\rho)=(1-s)\, \rho+s\, \sigma_1\rho\sigma_1$ and a phase flip channel $\mathcal{P}_t(\rho)=(1-t) \,\rho+t\, \sigma_3\rho\sigma_3$, corresponding to Pauli channels $\map E_{\vec p}$ and $\map F_{\vec q}$ with $\vec p  =  (1-s,s,0,0)$ and $\vec q  =  (1-t,0,0,t)$, respectively \cite{salek2018quantum}. For $s=t=1/2$, the two channels  are entanglement-breaking,  and therefore unable to transmit any quantum information. In contrast, the channel $\map C_-$ of Equation (\ref{C-}) is the unitary gate $\sigma_2$, and therefore it allows for the noiseless heralded transmission of a qubit, meaning that the receiver can decode the message without any error through the channel $\map C_-$.

The possibility of noiseless heralded  quantum communication is an important difference between the communication model with independent quantum channels in a superposition of orders and a related communication model with independent quantum  channels traversed in a superposition of paths \cite{gisin2005error,abbott2018communication,chiribella2019quantum}.  The transmission of quantum information through one of two channels $\map E$ and $\map F$ is described by a controlled-channel with Kraus operators \cite{oi2003interference,abbott2018communication,chiribella2019quantum}
\begin{align}\label{cohdef}
W'_{ij}   =      \beta_j  \,  E_i   \otimes  |0\>\<0|    +  \alpha_i \,  F_j   \otimes |1\>\<1|  \, ,
\end{align}
where $\alpha_i$ and $\beta_j$ are  complex amplitudes, and the states $|0\>$ and $|1\>$ represent paths of the information carrier, going through channels $\map E$ and $\map F$, respectively.  In this setting, Gisin {\em et al} showed that preparing the path in a coherent superposition leads to  a heralded reduction of the noise \cite{gisin2005error}. More recently, it was shown that the superposition of paths can also  increase the overall capacity, enabling deterministic communication through depolarizing channels \cite{abbott2018communication} and even complete erasure channels \cite{chiribella2019quantum}.

The communication enhancements due to superpositions of paths (\ref{cohdef}) and superposition of orders (\ref{switchdef2}) share several common features,  in particular  the crucial role of coherence between alternative configurations of the communication channels.      Nevertheless, they also exhibit interesting  differences.   One such difference concerns the possibility of noiseless heralded communication: while placing two independent noisy channels  in a superposition of orders can lead to heralded noiseless communication,  placing them on two alternative paths only lead to a partial noise reduction \cite{salek2018quantum,chiribella2018indefinite}.  Interestingly, this feature  is  balanced by the fact that communication enhancements due to superpositions of paths are more common, while the enhancements due to the superposition of order require  a  specific   matching between coherence in the superposition and commutativity of the channels \cite{loizeau2019channel}.

While every real experiment involves noise and imperfections, the in-principle  possibility of noiseless communication  through superposition of orders suggests that the experimental fidelities can be arbitrarily close to 1.  In our experiment, we pushed towards this target by adopting a phase-locked system  (described in Ref.~\cite{SM}) to ensure the stability of the path interferometer.   Thanks to  phase locking, we managed to obtain  an average fidelity of $0.9747 \pm 0.0012$ with the unitary gate  $\sigma_2$  predicted by the theory. In addition, we used the reconstructed channel matrix   $\gamma_{ij}$ to find the input state that maximises the coherent information in Eq.~(\ref{equ3}). The optimisation can be reduced to three real parameters $(\alpha,  \theta, \psi)$ by parameterising the qubit state $\rho$ as $\rho=\alpha_0  |\phi\rangle\langle \phi|+ (1-\alpha_0)|\phi_{\bot}\rangle\langle \phi_{\bot}|$, where  $|\phi\rangle=\cos(\theta)|0\rangle+\sin(\theta)\mathrm{e}^{i\psi}|1\rangle$ and $|\phi_{\bot}\rangle=\sin(\theta)|0\rangle-\cos(\theta)\mathrm{e}^{i\psi}|1\rangle$ are basis states. Since the entropy exchange is independent of the choice of purification, the parameters $(\alpha_0,  \theta,\psi)$ completely determine the coherent information.

\begin{figure}[tbph]
\includegraphics [width= 1\columnwidth]{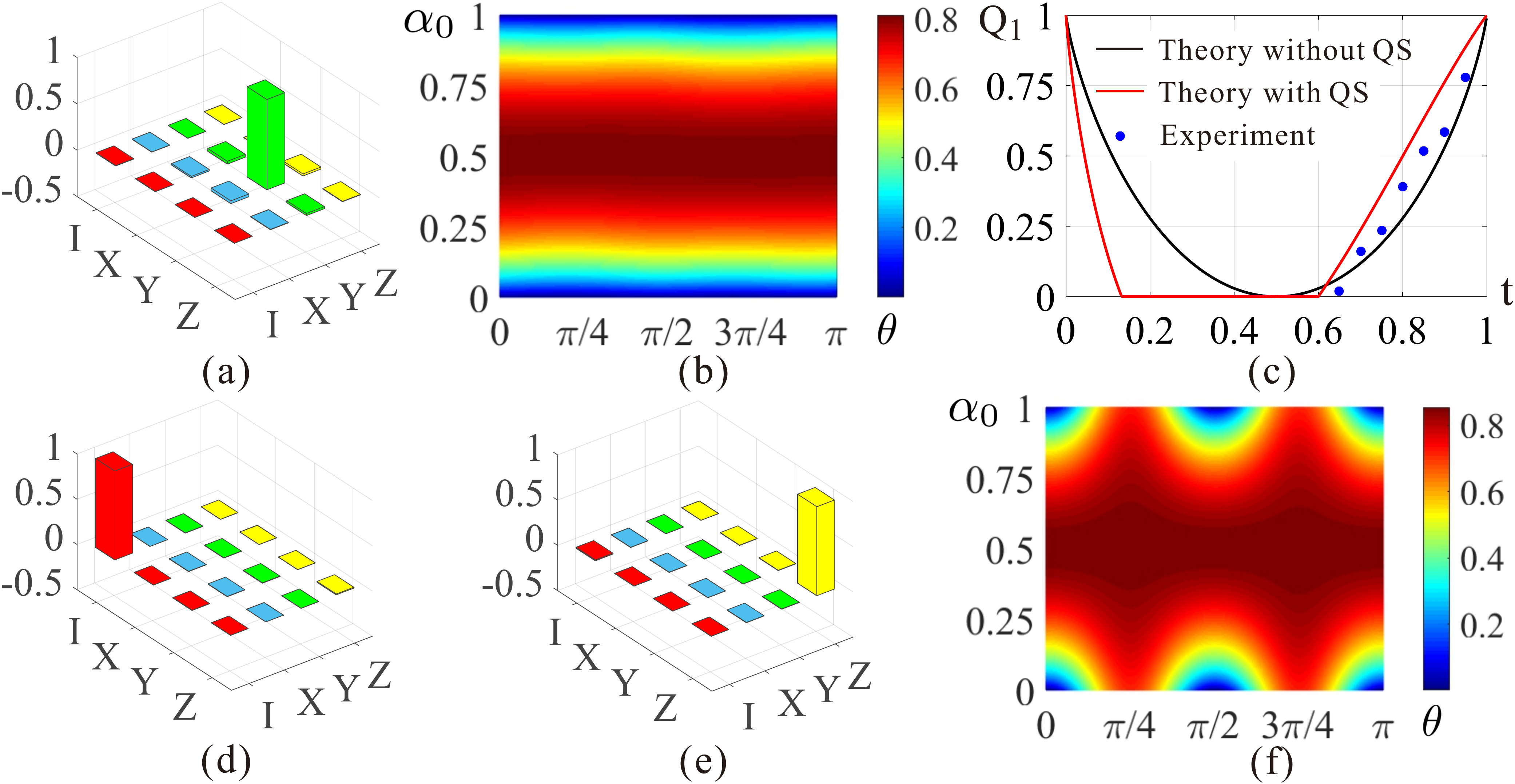}
\caption{Transmission of quantum information through entanglement-breaking channels in a superposition of causal orders.  (a) Real part of the reconstructed matrix of $\mathcal{S}(\mathcal{B}_s,\mathcal{P}_t)$, for $s=t=1/2$. (b) Coherent information $\mathcal{I}_c$ for the channel $\map C_-$ with $s=t=\frac{1}{2}$ as a function of  $\theta$ and $\alpha_0$, for fixed $\psi=0.790\pi$. (c) One-shot coherent information $Q_1$  for the channel $\mathcal{S}(\mathcal{B}_t,\mathcal{P}_t)$ with $t\in[0,1]$. The experimental results are marked as blue dots, while the correspond theoretical predictions are plotted in the red line. For comparison, the one-shot coherent information of the two dephasing channels combined in a definite order is also shown in the black line. (d-e) Real parts of the reconstructed matrices of $\map C_+$ and $\map C_-$ of the channel $\mathcal{S}(\mathcal{F},\mathcal{F})$. (f) Coherent information $\mathcal{I}_c$ for channel $\mathcal{S}(\mathcal{F},\mathcal{F})$ as a function of $\theta$ and $\alpha_0$, for fixed $\psi=0.155\pi$. Error bars are not visible in the figure as they are smaller than the marker size.
}
\label{fig:3}
\end{figure}

Fig.~\ref{fig:3} shows the experimental results for the coherent information $\mathcal{I}_c$ of the channel $\map C_-$ and of the whole channel $\mathcal{S}(\mathcal{B}_s,\mathcal{P}_t)$. For channel $\map C_-$, the result is a one-shot coherent information $Q_1$ of about $0.812\pm 0.003$, obtained with parameters $\alpha_0=0.500$, $\theta=0.0723\pi$, and $\psi=0.790\pi$.  The dependence of the coherent information $\mathcal{I}_c$ on $\alpha_0$ and $\theta$ is shown  in Fig.~\ref{fig:3} (b) for  fixed $\psi=0.790\pi$. More generally, $Q_1$ of the channel $\mathcal{S}(\mathcal{B}_s,\mathcal{P}_t)$ is further show in Fig.~\ref{fig:3} (c) as a function of $t$, with  $s=t$. One can find out that, as long as $t>0.62$, $Q_1$ of $\mathcal{S}(\mathcal{B}_t,\mathcal{P}_t)$ (red line) surpasses the coherent information when the two channels are combined in a fixed order (black line). Our results (blue dots) verified the existence of the advantage of indefinite causal order and the possibility of heralded, high-fidelity communication. The deviation of the channel capacities from their theoretical predictions are due to imperfect channel simulations and measurement error.

An even more radical example of transmission of quantum information  in a superposition of order corresponds to two entanglement-breaking channels   $\mathcal{F} (\rho)=1/2(\sigma_1\rho\sigma_1+\sigma_2\rho\sigma_2)$ \cite{chiribella2018indefinite}. In normal conditions, the channel $\map F$ cannot transmit any quantum information. Still, when two such channels are inserted in the QS, the channels $\map C_+$ and $\map C_-$ are both unitary, enabling the deterministic noiseless transmission of one qubit. This effect is more dramatic than the aforementioned heralded noiseless communication, as heralded noiseless transmission alone is not enough to guarantee a non-zero quantum capacity.

In our experiment,
we find  that the channels $\map C_+$ and $\map C_-$ have fidelities  $0.9823\pm0.0013$ and $0.9846\pm0.0014$ with the corresponding unitary gates (Fig.~\ref{fig:3} (d-e)). The one-shot coherent information $Q_1$ is $0.855\pm0.004$ and can be obtained when $\alpha_0=0.500$, $\theta=0.7575\pi$, and $\psi=0.155\pi$. Fig.~\ref{fig:3} (f) reports the result of coherent information $\mathcal{I}_c$ of channel $\mathcal{S}(\mathcal{F},\mathcal{F})$ varying with the parameters $\alpha_0$ and $\theta$, while $\psi$ is set to be $0.155\pi$.

\begin{figure}[tbph]
\begin{center}
\includegraphics [width=1\columnwidth]{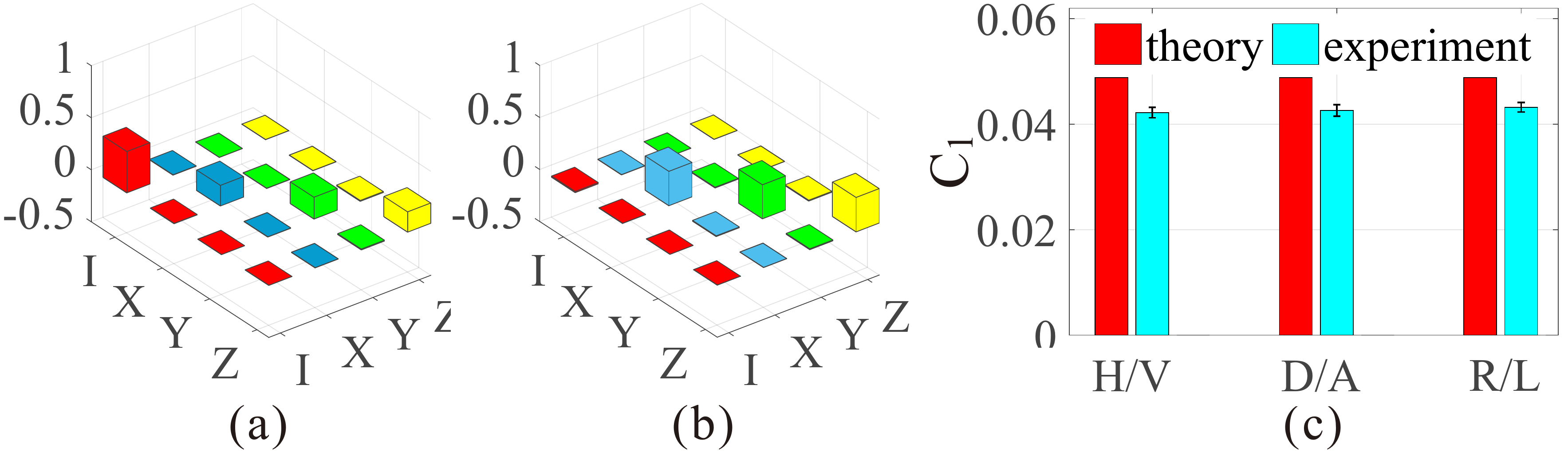}
\end{center}
\caption{Communication of classical information through  depolarizing channels. (a-b) Real parts of the matrices of $\map C_+$ and $\map C_-$ for the channel $\mathcal{S}(\mathcal{D},\mathcal{D})$. (c) Mutual information for $\mathcal{S}(\mathcal{D},\mathcal{D})$, when encoding and decoding with H/V, D/A, and R/L. Results from theory and experiments are represented by red and green bars respectively.
}
\label{fig:2}
\end{figure}

\paragraph{Transmitting classical information with depolarising  channels.---} Consider two completely depolarising channels $\mathcal{D}(\rho)=1/4\sum_{i=0}^3\sigma_i\rho\sigma_i$.  In this case, the resulting channels $\map C_+$ and $\map C_-$   can transmit classical information, despite the fact that no classical information can be sent through each depolarising channel individually \cite{ebler2018enhanced}.
This scenario demonstrates an advantage over all communication protocols where multiple depolarising channels are used in a sequence, and any intermediate operation between them does not transfer information from the internal degrees of freedom of the particle to its path \cite{chiribella2019quantum}.  This protocol was experimentally demonstrated in Ref. \cite{goswami2018communicating} using orbital angular momentum modes as the information carriers. Our  implementation uses polarisation states, which achieve a higher communication performance, allowing us  to demonstrate a non-zero communication capacity with more than 38.7 standard deviations.

Fig.~\ref{fig:2} (a-b) illustrate the real parts of $\map C_+$ and $\map C_-$ of reconstructed $\mathcal{S}(\mathcal{D},\mathcal{D})$ (see the imaginary parts in Ref.~\cite{SM}). The fidelities are $0.9989\pm 0.0001$ and $0.9903\pm 0.0017$, thereby establishing the high quality of our channel. To lower bound  the classical capacity, we chose three kinds of signal states, specifically, $\{|H\rangle, |V\rangle\}$, $\{|D\rangle, |A\rangle\}$, and $\{|R\rangle, |L\rangle \}$; the input distribution $p(x)$ is set to maximize the mutual information according to Eq.~(\ref{equ2}). The  mutual information with respect to these three choices of signal states is $0.0422\pm0.0010$, $0.0426\pm0.0011$, and $0.0432\pm0.0009$, matching well the theoretical value of $0.0488$ (Fig.~\ref{fig:2}(c)). The error bars are estimated by Monte Carlo simulations.

\paragraph {Conclusion.---} We have experimentally demonstrated the high-fidelity transmission of quantum information through communication channels in a coherent superposition of alternative orders.
Our experiments can be viewed as  simulations of exotic communication scenarios where the sender and the receiver are embedded in a  quantum  spacetime, and  the order between the noisy processes occurring in two different regions is indefinite.  At the same time, the accurate coherent control  over multiple communication lines, achieved in our setup, is a flexible primitive that can also be used to demonstrate more general communication advantages,  arising from  superpositions of paths \cite{gisin2005error,abbott2018communication,chiribella2019quantum}, directions of communication \cite{delsanto2018two}, or encoding operations \cite{guerin2018communication}.  Our results suggest that  coherent control over multiple communication channels may also have  applications to quantum communication in the  ordinary, classically well-defined spacetime.


\begin{acknowledgments}
[\emph{Note added.} During the preparation of this manuscript, the authors became aware of a work by K. Goswami et al~\cite{goswami2018communicating}. that independently demonstrated the enhanced effect of a superposition of causal orders on transmitting classical information.]

\paragraph {Acknowledgments.---}We thank Kang-Da Wu, Jin-Shi Xu, Xiao-Ye Xu, Chao Zhang, Kai Sun and Yong-Xiang Zheng for valuable discussions. This work was supported by the National Key Research and Development Program of China (No.\ 2017YFA0304100, 2016YFA0301300, 2016YFA0301700), NSFC (Nos.\ 11774335, 11504253, 11874345, 11821404, 11675136), the Key Research Program of Frontier Sciences, CAS (No.\ QYZDY-SSW-SLH003), the Fundamental Research Funds for the Central Universities,  the Anhui Initiative in Quantum Information Technologies (Nos.\ AHY020100, AHY060300), the John Templeton Foundation (grant 60609, Quantum
Causal Structures),  the Croucher Foundation, and the Hong Research Grant Council
through grants 17326616 and 17307719.  This publication was made possible through the support of a grant from the John Templeton Foundation. The opinions expressed in this publication are those of the authors and do not necessarily reflect the views of the John Templeton Foundation.
\end{acknowledgments}

\newpage
\setcounter{figure}{0}
\renewcommand{\thefigure}{S\arabic{figure}}
\setcounter{equation}{0}
\renewcommand{\theequation}{S\arabic{equation}}
\def\>{\rangle}
\def\<{\langle}
\def\bb{\langle\!\langle}
\def\kk{\rangle\!\rangle}


\onecolumngrid
\begin{center}
{\large{Supplementary Materials for \\
Experimental transmission of quantum information using a superposition of causal orders}}
\end{center}


{\em Phase locking effect.---} We used a phase locked system composed by piezoelectric transducer to overcome the phase drift of the Mach-Zenhder (MZ)  interferometer in the quantum switch caused by air turbulence. We observed, as shown in Fig.~\ref{fig:1}, the phase stability of the MZ interferometer for two hours, between which an extra relative phase of $\pi$ was added between the two spatial modes at the point of one hour. The overall visibility during this time was 0.979. In our experiment, to switch two Pauli channels, we need to perform $4\times4=16$ separate experiments to obtain all needed operations. To perform quantum process tomography of the switched channels, we need to prepare 6 kind of signal states, i.e. $|H\rangle$, $|V\rangle$, $|D\rangle$, $|A\rangle$, $|R\rangle$, and $|L\rangle$ and reconstruct all the out put states, which requires 3 ovservables $\sigma_1$, $\sigma_2$, and $\sigma_3$ to be measured. So, the total number of measurement settings would be $16\times6\times3=288$. As each setting took 4~s to collect the data and 3~s to set the angles of the half wave plates and quarter wave plates, the total measurement time was under 35~min. The visibility of our setup during this time was maintained at high quality.

\begin{figure*}[tbph]
\begin{center}
\includegraphics [width=0.5\columnwidth]{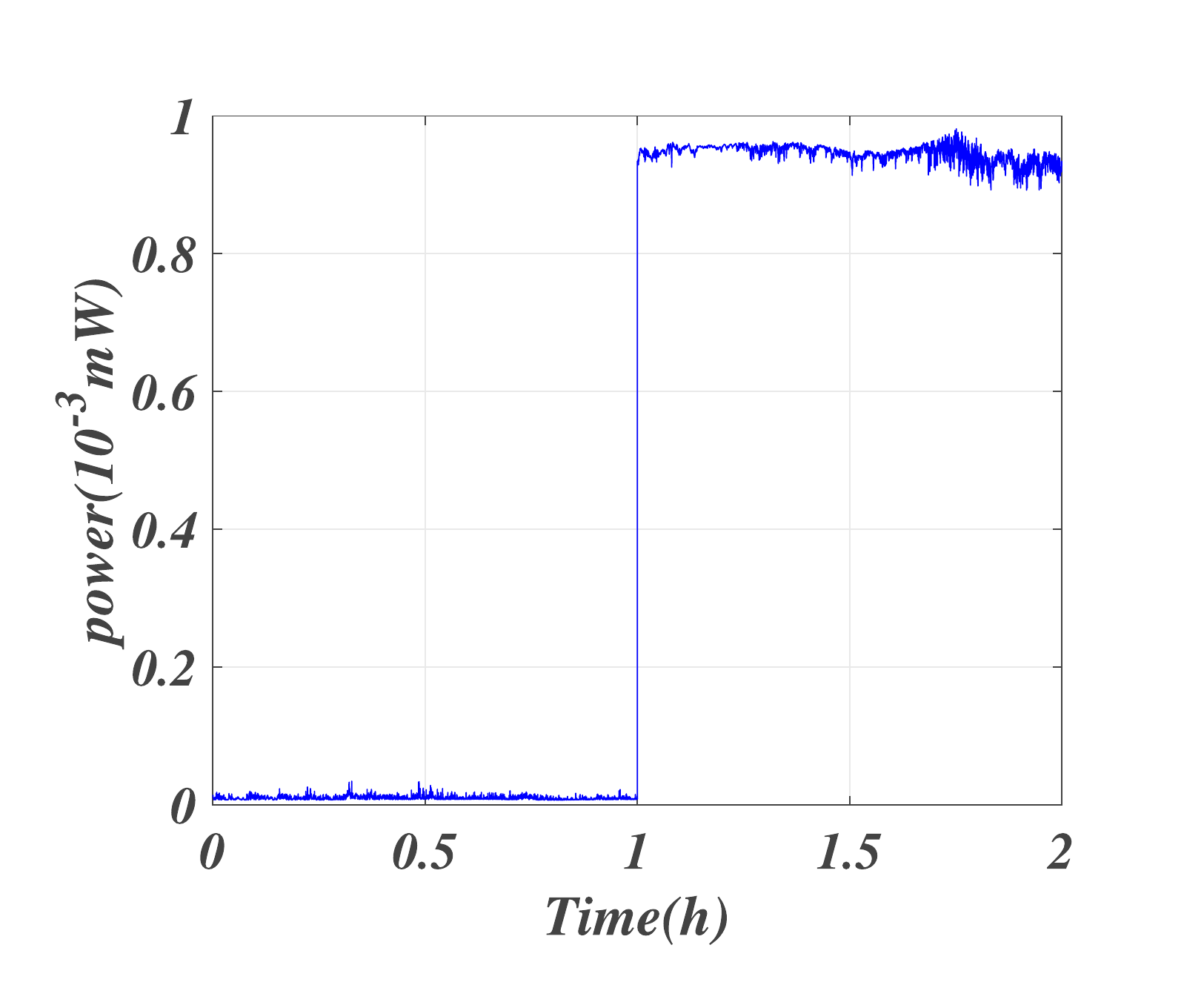}
\end{center}
\caption{Result of visibility of the MZ interferometer in two hours. At the point of 1 hour, an extra phase of $\pi$ was added between spatial mode 1 and spatial mode 2. The overall visibility remained 0.979 during this period.
}
\label{fig:1}
\end{figure*}

\bigskip
{\em Communication through switched Pauli channel.---} The Kraus representation of a generic Pauli channel is given by
\begin{equation}\label{equ1}
\begin{aligned}
\mathcal{E}_{\vec{p}}=\sum_{i=0}^3 p_i\sigma_i\rho\sigma_i,
\end{aligned}
\end{equation}
where Pauli operators $\sigma_i$ are in the set $\{I,X,Y,Z\}$ and $\vec{p}=(p_0,p_1,p_1,p_3)$ denotes $\sigma_i$'s probability vector. When two Pauli channels with probability vectors of $\vec{p}=(p_0,p_1,p_1,p_3)$ and $\vec{q}=(q_0,q_1,q_1,q_3)$ assembled in a quantum switch, the new channel, noted as $\mathcal{S}(\mathcal{E}_{\vec{p}},\mathcal{E}_{\vec{q}})(\rho)$ here, can be defined as
\begin{equation}\label{equ2}
\mathcal{S}(\mathcal{E}_{\vec{p}},\mathcal{E}_{\vec{q}})(\rho)=\sum_{i,j=0}^4W_{ij}(\rho\otimes|c\rangle\langle c|)W_{ij}^\dagger,
\end{equation}
with its Kraus operators
\begin{equation}\label{equ3}
W_{ij}=\sqrt{p_iq_j}\sigma_i\sigma_j\otimes|0\rangle\langle0|_c+\sqrt{q_jp_i}\sigma_j\sigma_i\otimes|1\rangle\langle1|_c.
\end{equation}
From Eq.~\ref{equ1} and Eq.~\ref{equ2}, the output state of the channel \begin{align}\label{switched}
 \mathcal{S}(\mathcal{E}_{\vec{p}},\mathcal{E}_{\vec{q}})(\rho)= r_+\map C_+(\rho)\otimes|+\rangle\langle+|+r_-\map C_-(\rho)\otimes|-\rangle\langle-| \, ,
 \end{align}
 where  $r_+$ and $r_-$ are the probabilities defined by  $r_- : =r_{12}  +  r_{23}  + r_{13}$,  $r_{ij}:  =   p_iq_j  + p_j q_i$, and $r_+ : = 1-  r_-$ are probabilities, and $\map C_+$ and $\map C_-$ are the Pauli channels defined by
 \begin{align}
 \map C_+=\frac{(\sum_{i=0}^3  r_{ii}/2)\rho+\sum_{i=1}^3 \,  r_{0i} \, \sigma_i\rho\sigma_i}{\sum_i  \,  (r_{0i}+  r_{ii}/2)}
 \end{align}
 and
 \begin{align}
 \map C_-=\frac{[   r_{12} \,  \sigma_3\rho\sigma_3+  r_{23}  \,\sigma_1\rho\sigma_1+  r_{31} \,   \sigma_2\rho\sigma_2]}{  r_{12} + r_{23} +  r_{13}} \, .
 \end{align}
If a bit flip channel ($\mathcal{B}_s(\rho)=(1-s)\rho+s\sigma_1\rho\sigma_1$) and a phase flip channel ($\mathcal{P}_t(\rho)=(1-t)\rho+t\sigma_3\rho\sigma_3$) are used, the output state is
\begin{eqnarray}\label{switchedDephasing}
\mathcal{S}(\mathcal{B}_s,\mathcal{P}_t)(\rho)&=&((1-t-s+2ts)\rho+s(1-t)\sigma_1\rho\sigma_1+(1-s)t\sigma_3\rho\sigma_3)\otimes|+\rangle\langle+|_c+(ts\sigma_2\rho\sigma_2)\otimes|-\rangle\langle-|_c,\\ r_-&=&s+t-2ts,\qquad r_+=1-s-t+2ts.  \, .
\end{eqnarray}
If two entanglement-breaking channels $\mathcal{F}(\rho)=1/2(\sigma_1\rho\sigma_1+\sigma_2\rho\sigma_2)$  are used, the output state is
\begin{eqnarray}\label{switchedEB}
\mathcal{S}(\mathcal{F},\mathcal{F})(\rho)&=&\rho/2\otimes|+\rangle\langle+|_c+\sigma_3\rho\sigma_3/2\otimes|-\rangle\langle-|_c,\\ r_-&=&\frac{1}{2},\qquad r_+=\frac{1}{2} \, .
\end{eqnarray}
If two depolarizing channels $\mathcal{D}(\rho)=1/4\sum_{i=0}^3\sigma_i\rho\sigma_i$ are used, the output state is \begin{eqnarray}\label{switchedD}
\mathcal{S}(\mathcal{D},\mathcal{D})(\rho)&=&(I/4+\rho/8)\otimes|+\rangle\langle+|_c+(I/4-\rho/8)\otimes|-\rangle\langle-|_c,\\ r_-&=&\frac{3}{8},\qquad r_+=\frac{5}{8} \, .
\end{eqnarray}

\begin{figure*}[tbph]
\begin{center}
\includegraphics [width=0.73\columnwidth]{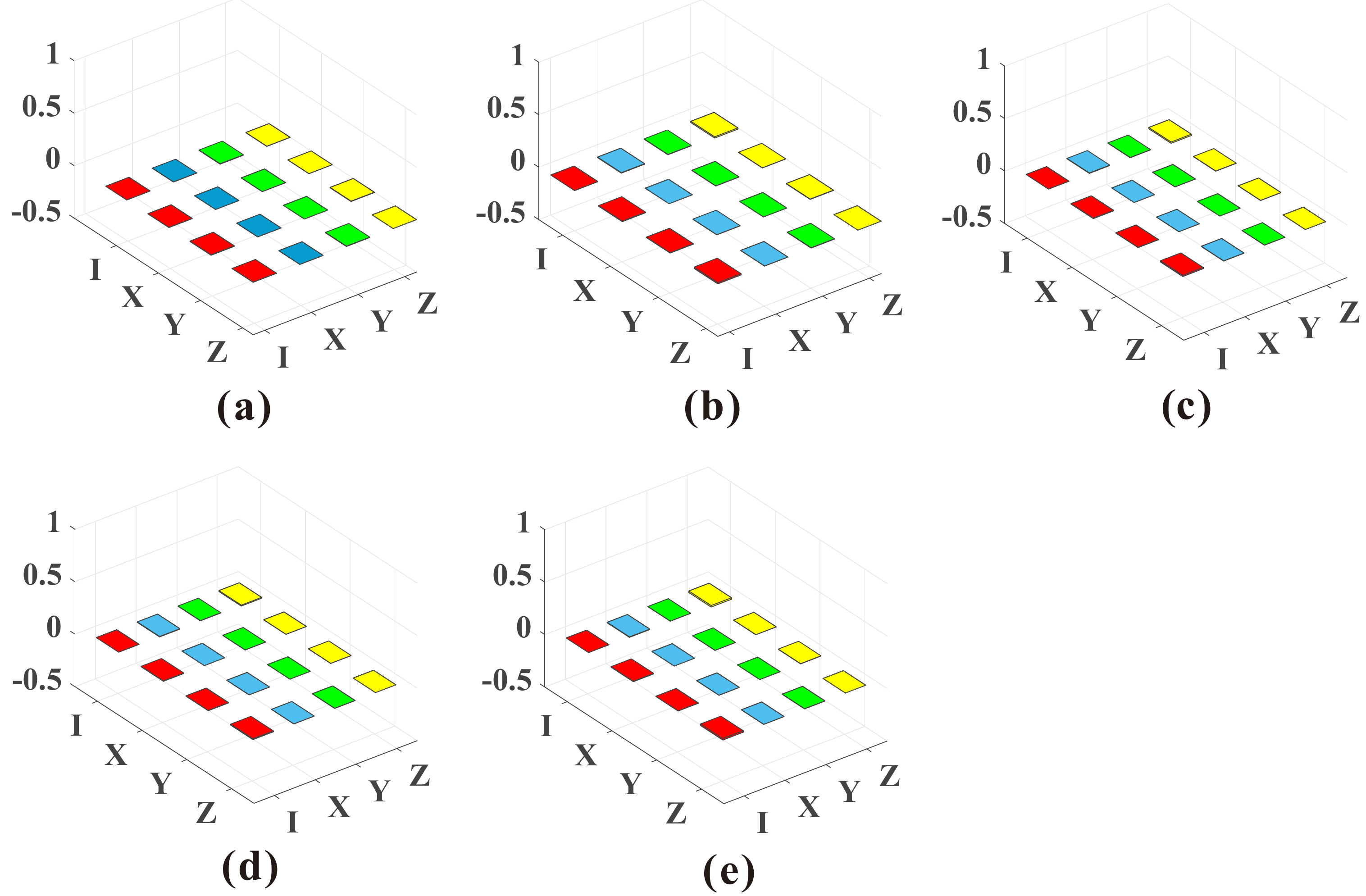}
\end{center}
\caption{The imaginary parts of the matrices of the channels: (a-b) for $\map C_+$ and $\map C_-$ of switched depolarizing channel $\mathcal{S}(\mathcal{D},\mathcal{D})$, (c) for $\map C_-$ of switched dephasing channel $\mathcal{S}(\mathcal{B}_s,\mathcal{P}_t)$ with $s=t=1/2$, (d-e) for $\map C_+$ and $\map C_-$ of switched entanglement-breaking channel $\mathcal{S}(\mathcal{F},\mathcal{F})$.
}
\label{fig:2}
\end{figure*}

\bigskip
{\em Further information on the experimental results.---} Fig.~\ref{fig:2} shows the imaginary parts of the matrices of the channels: Fig.~\ref{fig:2}(a-b) for $\map C_+$ and $\map C_-$ of switched depolarizing channel $\mathcal{S}(\mathcal{D},\mathcal{D})$, Fig.~\ref{fig:2}(c) for $\map C_-$ of switched dephasing channel $\mathcal{S}(\mathcal{B}_s,\mathcal{P}_t)$ with $s=t=1/2$, Fig.~\ref{fig:2}(d-e) for $\map C_+$ and $\map C_-$ of switched entanglement-breaking channel $\mathcal{S}(\mathcal{F},\mathcal{F})$.

\end{document}